\documentclass[article,preprint,groupedaddress]{revtex4}
\usepackage{epsfig}
\usepackage{graphicx}
\usepackage{amssymb}

\begin{document}
\title{Charged reflecting shells supporting non-minimally coupled massless scalar field configurations}
\author{Shahar Hod}
\affiliation{The Ruppin Academic Center, Emeq Hefer 40250, Israel}
\affiliation{ }
\affiliation{The Hadassah Institute, Jerusalem 91010, Israel}
\date{\today}

\begin{abstract}
\ \ \ We study {\it analytically} the physical and mathematical
properties of spatially regular massless scalar field configurations
which are non-minimally coupled to the electromagnetic field of a
spherically symmetric charged reflecting shell. In particular, the
Klein-Gordon wave equation for the composed
charged-reflecting-shell-nonminimally-coupled-linearized-massless-scalar-field
system is solved analytically. Interestingly, we explicitly prove
that the discrete resonance spectrum
$\{R_{\text{s}}(Q,\alpha,l;n)\}^{n=\infty}_{n=1}$ of charged shell
radii that can support the non-minimally coupled massless scalar
fields can be expressed in a remarkably compact form in terms of the
characteristic zeros of the Bessel function (here $Q$, $\alpha$, and
$l$ are respectively the electric charge of the central supporting
shell, the dimensionless non-minimal coupling parameter of the
Maxwell-scalar theory, and the angular harmonic index of the
supported scalar configuration).
\end{abstract}
\bigskip
\maketitle

\section{Introduction}

Black-hole spacetimes are characterized by the presence of event
horizons, well defined boundaries on which matter and radiation
fields are characterized by purely ingoing (absorbing) boundary
conditions. This remarkable property of classical black-hole
horizons has led Wheeler \cite{Whee,Car,Bektod} to conjecture that
static field configurations cannot be supported in the exterior
regions of black-hole spacetimes with spatially regular horizons.

The {\it absorbing} boundary conditions, which characterize the
behavior of matter fields on classical black-hole horizons, have
played a key role in various no-hair theorems
\cite{Bek1,Her1,Hodstationary,BekMay,Hod1} that explicitly proved,
in accord with Wheeler's conjecture \cite{Whee,Car,Bektod}, that
spatially regular matter configurations made of minimally coupled
scalar fields (or scalar fields with a non-minimal coupling to the
Ricci curvature scalar) cannot be supported in black-hole
spacetimes.

Intriguingly, recent studies \cite{Hodrec,Hodnww,Sar} have revealed
the fact that absorbing boundary conditions are actually not a
necessary condition for a no-hair property of compact physical
objects. In particular, the no-hair theorems presented in
\cite{Hodrec,Hodnww} have explicitly proved that spherically
symmetric horizonless compact stars with {\it reflecting} (rather
than absorbing) boundary conditions cannot support spatially regular
scalar field configurations.

Interestingly, it has recently been proved in the physically
important works \cite{Herch,Herr} (see also \cite{Hodwk}) that the
no-hair conjecture can be violated in composed
Einstein-Maxwell-scalar theories in which the scalar fields are
non-minimally coupled to the Maxwell electromagnetic tensor. In
particular, it has been explicitly demonstrated in \cite{Herch,Herr}
that spherically symmetric charged black holes can support massless
scalar field configurations which are non-minimally coupled to the
electromagnetic field of the central supporting charged black hole
\cite{Notenrf,Nwr1,Nwr2,Nwr3}.

In the present compact paper we shall explicitly prove that,
similarly to the familiar case of charged black-hole spacetimes with
{\it absorbing} boundary conditions, charged compact shells with
{\it reflecting} boundary conditions can also support spatially
regular massless scalar field configurations which are characterized
by a non-minimal coupling to the electromagnetic field of the
central supporting shell.

In particular, using {\it analytical} techniques \cite{Noteap,Peng},
we shall explicitly prove below that, for a spherically symmetric
compact reflecting shell of electric charge $Q$, there exists an
infinitely large discrete resonance spectrum
$\{R_{\text{s}}(Q,\alpha;n)\}^{n=\infty}_{n=1}$ of shell radii that
can support non-minimally coupled massless scalar field
configurations [here the physical parameter $\alpha$ is the
dimensionless coupling parameter between the supported scalar field
and the electromagnetic field of the central charged shell, see Eq.
(\ref{Eq3}) below].

\section{Description of the system}

We shall analyze the physical and mathematical properties of
massless scalar field configurations which are non-minimally coupled
to the electromagnetic field of charged reflecting shells.

The spatial behavior of the non-minimally coupled scalar field
configurations is determined by the differential equation
\cite{Herch,Herr,Notelag,Noteunits}
\begin{equation}\label{Eq1}
\nabla^\nu\nabla_{\nu}\phi={1\over4}f_{,\phi}{\cal I}\  ,
\end{equation}
where the source term ${\cal I}$, which governs the non-trivial
coupling between the supported scalar field $\phi$ and the
electromagnetic Maxwell tensor $F_{\mu\nu}$ of the central charged
shell, is given by the simple expression \cite{Noteiq}
\begin{equation}\label{Eq2}
{\cal I}=F_{\mu\nu}F^{\mu\nu}\  .
\end{equation}
In the weak-field regime, the scalar coupling function $f(\phi)$ is
assumed to be characterized by the quadratic behavior
\cite{Herch,Herr,Hodwk}
\begin{equation}\label{Eq3}
f(\phi)=1+\alpha\phi^2\  ,
\end{equation}
where the dimensionless physical parameter $\alpha$ \cite{Notealp0}
couples the scalar field to the electromagnetic field of the charged
supporting shell \cite{Notesoln,soln}.

In order to facilitate a fully {\it analytical} treatment of the
composed charged-shell-nonminimally-coupled-massless-scalar-field
system, we shall work within the flat-space approximation, in which
case the scalar equation (\ref{Eq1}) can be written in the compact
mathematical form
\begin{equation}\label{Eq4}
{{d}\over{dr}}\Big(r^2{{dR_{lm}}\over{dr}}\Big)-\Big[l(l+1)-{{\alpha
Q^2}\over{r^2}}\Big]R_{lm}=0\  ,
\end{equation}
where $Q$ is the electric charge of the central supporting shell
\cite{NoteQQ}. The radial scalar eigenfunction $R_{lm}(r)$ is
defined by the relation
\begin{equation}\label{Eq5}
\phi(r,\theta,\phi)=\sum_{lm}R_{lm}(r)Y_{lm}(\theta)e^{im\phi}\ .
\end{equation}
Here the integer parameters $\{l,m\}$ are the angular harmonic
indices of the supported scalar field modes [note that the
characteristic eigenvalue of the angular scalar eigenfunction
$Y_{lm}(\theta)$ is given by the simple expression $l(l+1)$].

Below we shall consider spatially regular non-minimally coupled
scalar field configurations whose radial profiles decay
asymptotically \cite{Herch,Herr},
\begin{equation}\label{Eq6}
\phi(r\to\infty)\to 0\  .
\end{equation}
In addition, we shall assume that the charged shell, which supports
the non-minimally coupled scalar field configurations, is
characterized by a reflecting surface of radius $R_{\text{s}}$ with
Dirichlet boundary conditions. This property of the central
supporting shell dictates the inner reflecting boundary condition
\begin{equation}\label{Eq7}
\phi(r=R_{\text{s}})=0\
\end{equation}
for the composed
charged-shell-nonminimally-coupled-massless-scalar-field
configurations.

As we shall now prove, the radial differential equation (\ref{Eq4})
for the supported scalar field configurations, supplemented by the
boundary conditions (\ref{Eq6}) and (\ref{Eq7}), determines an
infinitely large discrete resonance spectrum of radii,
$\{R_{\text{s}}(Q,\alpha,l;n)\}^{n=\infty}_{n=1}$, which, for given
values $\{Q,\alpha,l\}$ of the physical parameters of the system,
characterize the central reflecting shells that can support the
non-minimally coupled static scalar field configurations. In
particular, in the next section we shall explicitly demonstrate that
the physical and mathematical properties of the composed
charged-reflecting-shell-nonminimally-coupled-massless-scalar-field
configurations can be studied {\it analytically}.

\section{The resonance spectrum of the composed
charged-reflecting-shell-nonminimally-coupled-massless-scalar-field
configurations}

In the present section we shall determine analytically the discrete
resonance spectrum $\{R_{\text{s}}(Q,\alpha,l;n)\}^{n=\infty}_{n=1}$
which characterizes the charged reflecting shells that can support
the spatially regular static configurations of the non-minimally
coupled massless scalar fields. This to this, it proves useful to
define the radial scalar eigenfunction
\begin{equation}\label{Eq8}
\psi_{lm}\equiv rR_{lm}\  ,
\end{equation}
in terms of which the radial equation (\ref{Eq4}) can be written in
the form
\begin{equation}\label{Eq9}
{{d^2\psi_l}\over{dr^2}}-V_l\psi=0\
\end{equation}
of a Schr\"odinger-like ordinary differential equation. Here
\cite{Herch,Herr}
\begin{equation}\label{Eq10}
V_l\equiv V(r;l,\alpha)={{l(l+1)}\over{r^2}} -{{\alpha
Q^2}\over{r^4}}\
\end{equation}
is the effective radial potential of the composed
charged-shell-scalar-field system \cite{Notenlm}.

Interestingly, as we shall now show, the Schr\"odinger-like radial
equation (\ref{Eq9}), which determines the spatial behavior of the
static non-minimally coupled scalar configurations, is amenable to
an analytical treatment. In particular, the general mathematical
solution of this radial differential equation can be expressed in
terms of the Bessel function of the first kind (see Eq. 9.1.53 of
\cite{Abram}):
\begin{equation}\label{Eq11}
\psi(r)=A\cdot r^{1\over
2}J_{l+{1\over2}}\Big({{\sqrt{\alpha}Q}\over{r}}\Big)+B\cdot
r^{1\over 2}J_{-(l+{1\over2})}\Big({{\sqrt{\alpha}Q}\over{r}}\Big)\
,
\end{equation}
where $A$ and $B$ are normalization constants.

Using the characteristic small-$x$ behavior (see Eq. 9.1.10 of
\cite{Abram})
\begin{equation}\label{Eq12}
J_{\nu}(x\to0)={{({1\over2}x)^{\nu}}\over{\Gamma(\nu+1)}}\cdot[1+O(x^2)]\
\end{equation}
of the Bessel functions, one finds that the asymptotic large-$r$
($\sqrt{\alpha}Q/r\to 0$) behavior of Eq. (\ref{Eq11}) is given by
\begin{equation}\label{Eq13}
\psi(r\to\infty)=A[\Gamma(l+{3\over2})]^{-1}\cdot\Big({{\sqrt{\alpha}Q}\over{2}}\Big)^{l+{1\over2}}r^{-l}+
B[\Gamma(-l+{1\over2})]^{-1}\cdot\Big({{\sqrt{\alpha}Q}\over{2}}\Big)^{-(l+{1\over2})}r^{l+1}\
.
\end{equation}
Taking cognizance of the asymptotic boundary condition (\ref{Eq6}),
one deduces the simple relation $B=0$. We therefore find that the
spatially regular static configurations of the non-minimally coupled
massless scalar fields are characterized by the radial functional
behavior
\begin{equation}\label{Eq14}
\psi(r)=A\cdot r^{1\over
2}J_{l+{1\over2}}\Big({{\sqrt{\alpha}Q}\over{r}}\Big)\ .
\end{equation}

Taking cognizance of the inner boundary condition (\ref{Eq7}), which
characterizes the behavior of the supported scalar fields on the
surface of the central reflecting shell, one finds that the composed
charged-reflecting-shell-nonminimally-coupled-massless-scalar-field
system is characterized by the {\it discrete} resonance spectrum
\begin{equation}\label{Eq15}
R_{\text{s}}(Q,\alpha,l;n)={{\sqrt{\alpha}Q}\over{j_{l+{1\over2},n}}}\
\ \ ; \ \ \ n=1,2,3,...\
\end{equation}
of the supporting shell radii. Here $n=1,2,3,...$ is the resonance
parameter of the system and $j_{l+{1\over2},n}$ is the $n$th
positive zero of the Bessel function $J_{l+{1\over2}}(x)$. The real
zeros of the Bessel function were studied by many authors, see e.g.
\cite{Abram,Bes}.

\section{Neumann boundary conditions}

In the previous section we have analyzed the physical properties of
non-minimally coupled scalar field configurations which are
supported by a charged reflecting shell with reflecting Dirichlet
boundary conditions [see Eq. (\ref{Eq7})]. It is also physically
interesting to study the case of Neumann boundary conditions
\begin{equation}\label{Eq16}
{{d\phi}\over{dr}}=0\ \ \ \text{for}\ \ \ r=R_{\text{s}}\
\end{equation}
at the surface of the supporting shell.

Taking cognizance of Eqs. (\ref{Eq5}), (\ref{Eq8}), and
(\ref{Eq14}), one can write the inner boundary condition
(\ref{Eq16}) in the form
\begin{equation}\label{Eq17}
{{d\Big[r^{-{1\over
2}}J_{l+{1\over2}}\Big({{\sqrt{\alpha}Q}\over{r}}\Big)\Big]}\over{dr}}=0\
\ \ \text{for}\ \ \ r=R_{\text{s}}\  .
\end{equation}
Using Eq. 9.1.31 of \cite{Abram}, one can express Eq. (\ref{Eq17})
in the form
\begin{equation}\label{Eq18}
J_{l+{1\over2}}\Big({{\sqrt{\alpha}Q}\over{R_{\text{s}}}}\Big)+
{{\sqrt{\alpha}Q}\over{R_{\text{s}}}}\Big[J_{l-{1\over2}}\Big({{\sqrt{\alpha}Q}\over{R_{\text{s}}}}\Big)-
J_{l+{3\over2}}\Big({{\sqrt{\alpha}Q}\over{R_{\text{s}}}}\Big)\Big]=0\
.
\end{equation}

As we shall now show explicitly, the rather cumbersome resonance
equation (\ref{Eq18}) can be solved analytically in the regime
\begin{equation}\label{Eq19}
{{\sqrt{\alpha}Q}\over{R_{\text{s}}}}\gg1\
\end{equation}
of small shell radii. In particular, using Eq. 9.2.1 of
\cite{Abram}, one can write (\ref{Eq18}) in the remarkably compact
form
\begin{equation}\label{Eq20}
\cos\Big({{\sqrt{\alpha}Q}\over{R_{\text{s}}}}-{1\over2}l\pi\Big)+
O\Big[\Big({{\sqrt{\alpha}Q}\over{R_{\text{s}}}}\Big)^{-1}\Big]=0\ .
\end{equation}

From the resonance condition (\ref{Eq20}), one finds that the
composed
charged-reflecting-shell-nonminimally-coupled-massless-scalar-field
system with reflecting Neumann boundary conditions is characterized
by the resonance spectrum
\begin{equation}\label{Eq21}
R_{\text{s}}(Q,\alpha,l;n)={{\sqrt{\alpha}Q}\over{\big(n-{1\over2}+{1\over2}l\big)\pi}}\
\ \ ; \ \ \ n=1,2,3,...\
\end{equation}
in the regime (\ref{Eq19}) of small shell radii \cite{Notellnn}.

\section{Summary and Discussion}

Classical black holes with {\it absorbing} boundary conditions
cannot support minimally coupled scalar field configurations
\cite{Bek1,Her1,Hodstationary,BekMay,Hod1}. Interestingly, this
remarkable property is also shared by compact {\it reflecting} stars
with repulsive (rather than absorbing) boundary conditions.

Intriguingly, the recently published works \cite{Herch,Herr} have
revealed the physically important fact that charged black-hole
spacetimes with regular event horizons can support massless scalar
field configurations which are non-minimally coupled to the Maxwell
tensor of the charged spacetime.

In the present compact paper we have explicitly proved that charged
reflecting shells in flat spacetimes, like charged absorbing black
holes, can support spatially regular configurations of massless
scalar fields which are non-minimally coupled to the Maxwell tensor
of the central compact shell. In particular, using {\it analytical}
techniques, we have derived the remarkably compact dimensionless
formula [see Eq. (\ref{Eq15})] \cite{Notecd}
\begin{equation}\label{Eq22}
\alpha(R_{\text{s}},Q,l;n)=\Big({{R_{\text{s}}}\over{Q}}\Big)^2\times
j^2_{l+{1\over2},n}\ \ \ ; \ \ \ n=1,2,3,...\
\end{equation}
for the discrete resonance spectrum that characterizes the physical
coupling parameter $\alpha$ of the non-trivially coupled
Maxwell-scalar theory \cite{Notealp}.

The analytically derived resonance spectrum (\ref{Eq22}) implies
that, for given values $\{R_{\text{s}},Q\}$ of the physical
parameters of the central supporting shell, the dimensionless
coupling parameter $\alpha$ of the composed
charged-shell-nonminimally-coupled-massless-scalar-field theory is
an increasing function of the resonance parameter $n$. In
particular, in the regime $n\gg l$ of large overtone numbers, one
may use the asymptotic relation (see Eqs. 9.5.12 of \cite{Abram})
$j_{l+{1\over2},n}=\pi[n+{1\over2}l+O(n^{-1})]$ for the zeros of the
Bessel function, which yields the asymptotic large-$n$ behavior
\begin{equation}\label{Eq23}
\alpha(n\gg l)=\Big({{\pi R_{\text{s}}}\over{Q}}\Big)^2\times
\big(n+{1\over2}l\big)^2\
\end{equation}
of the resonance spectrum.

In addition, the resonance spectrum (\ref{Eq22}) implies that the
dimensionless coupling parameter $\alpha$ of the composed
charged-shell-nonminimally-coupled-massless-scalar-field
configurations is an increasing function of the angular harmonic
index $l$. In particular, in the regime $l\gg n$ of large angular
harmonic indices, one may use the asymptotic relation (see Eq.
9.5.14 of \cite{Abram})
$j_{l+{1\over2},n}=(l+{1\over2})[1+O(l^{-2/3})]$ for the zeros of
the Bessel function, which yields the asymptotic large-$l$ behavior
\cite{Notedls}
\begin{equation}\label{Eq24}
\alpha(l\gg n)=\Big({{R_{\text{s}}}\over{Q}}\Big)^2\times l^2\
\end{equation}
of the resonance spectrum.

\bigskip
\noindent
{\bf ACKNOWLEDGMENTS}
\bigskip

This research is supported by the Carmel Science Foundation. I would
like to thank Yael Oren, Arbel M. Ongo, Ayelet B. Lata, and Alona B.
Tea for helpful discussions.



\begin{thebibliography}{99}

\bibitem{Whee} R. Ruffini and J. A. Wheeler, Phys. Today {\bf 24}, 30
(1971).

\bibitem{Car} B. Carter, in {\it Black Holes}, Proceedings of 1972 Session of Ecole d'ete de Physique Theorique,
edited by C. De Witt and B. S. De Witt (Gordon and Breach, New York,
1973).

\bibitem{Bektod} J. D. Bekenstein, Physics Today {\bf 33}, 24 (1980).

\bibitem{Bek1} J. D. Bekenstein, Phys. Rev. D {\bf 5}, 1239 (1972).

\bibitem{Her1} C. A. R. Herdeiro and E. Radu, Int. J. Mod. Phys. D {\bf 24}, 1542014
(2015).

\bibitem{Hodstationary} S. Hod, Phys. Lett. B {\bf 713}, 505 (2012);
S. Hod, Phys. Lett. B {\bf 718}, 1489 (2013) [arXiv:1304.6474]; S.
Hod, Phys. Rev. D {\bf 91}, 044047 (2015) [arXiv:1504.00009].

\bibitem{BekMay} A. E. Mayo and J. D. Bekenstein and, Phys. Rev. D {\bf 54}, 5059 (1996).

\bibitem{Hod1} S. Hod, Phys. Lett. B {\bf 771}, 521 (2017) [arXiv:1911.08371];
S. Hod, Phys. Rev. D {\bf 96}, 124037 (2017) [arXiv:2002.05903].

\bibitem{Hodrec} S. Hod, Phys. Rev. D {\bf 94}, 104073 (2016) [arXiv:1612.04823].

\bibitem{Hodnww} S. Hod, Phys. Rev. D {\bf 96}, 024019 (2017) [arXiv:1709.01933];
S. Hod, Phys. Lett. B {\bf 773}, 208 (2017).

\bibitem{Sar} S. Bhattacharjee and S. Sarkar, Phys. Rev. D {\bf 95}, 084027 (2017).

\bibitem{Herch} C. A. R. Herdeiro, E. Radu, N. Sanchis-Gual, and J. A. Font,
Phys. Rev. Lett. {\bf 121}, 101102 (2018).

\bibitem{Herr} P. G. S. Fernandes, C. A. R. Herdeiro, A. M. Pombo, E. Radu, and N.
Sanchis-Gual, Class. Quant. Grav. {\bf 36}, 134002 (2019)
[arXiv:1902.05079].

\bibitem{Hodwk} S. Hod, Phys. Lett. B {\bf 798}, 135025 (2019) [arXiv:2002.01948].

\bibitem{Notenrf} See \cite{Nwr1,Nwr2,Nwr3} for the interesting case
of scalar fields which are non-minimally coupled to the Gauss-Bonnet
invariant and supported by black holes and reflecting shells.

\bibitem{Nwr1} D.D. Doneva and S. S. Yazadjiev, Phys. Rev. Lett. {\bf 120}, 131103 (2018).

\bibitem{Nwr2} H. O. Silva, J. Sakstein, L. Gualtieri, T.P. Sotiriou, and E. Berti,
Phys. Rev. Lett. {\bf 120}, 131104 (2018).

\bibitem{Nwr3} Y. Peng, Eur. Phys. J. C {\bf 80}, 202 (2020).

\bibitem{Noteap} The case of massive non-minimally coupled scalar
fields was studied {\it numerically} in the interesting work
\cite{Peng}. In the present paper we shall explicitly prove that the
case of massless non-minimally coupled scalar fields can be studied
{\it analytically}.

\bibitem{Peng} Y. Peng, Phys. Lett. B {\bf 804}, 135372 (2020).

\bibitem{Notelag} The scalar field equation (\ref{Eq1}) follows from the
Lagrangian density
$2\nabla_{\alpha}\phi\nabla^{\alpha}\phi+f(\phi){\cal I}$
\cite{Herch,Herr}. In addition, in order to facilitate a fully {\it
analytical} treatment of the composed
charged-shell-nonminimally-coupled-massless-scalar-field system, we
shall work within the flat-space approximation, in which case the
metric is described by the spherically symmetric line element
$ds^2=-dt^2+dr^2+r^2(d\theta^2+\sin^2\theta d\phi^2)$.

\bibitem{Noteunits} We shall use natural units in which $\hbar=c=1$.

\bibitem{Noteiq} Note that ${\cal I}=Q^2/r^4$ in the weak-field
regime.

\bibitem{Notealp0} We shall henceforth assume that $\alpha>0$.

\bibitem{Notesoln} It is important to emphasize the physically interesting fact
that, as nicely shown in \cite{soln}, this class of models with a
different coupling function $f(\phi)$ also allows for the existence
of everywhere regular (solitonic) field configurations.

\bibitem{soln} C. A. R. Herdeiro, J. M. S. Oliveira, and E. Radu, arXiv:1910.11021.

\bibitem{NoteQQ} Without loss of generality, we shall assume $Q>0$ for the electric charge of the compact
shell.

\bibitem{Notenlm} For brevity, we shall henceforth omit the angular harmonic parameters $l$ and $m$.

\bibitem{Abram} M. Abramowitz and I. A. Stegun, {\it Handbook of
Mathematical Functions} (Dover Publications, New York, 1970).

\bibitem{Bes} K. T. Tang, {\it Mathematical methods for engineers
and scientists3: Fourier analysis, partial differential equations
and variational models} (Springer, New York, 2006).

\bibitem{Notellnn} Note that the regime (\ref{Eq19}) of small shell radii
corresponds to large-$n$ values and/or lareg-$l$ values.

\bibitem{Notecd} The analytically derived resonance formula
(\ref{Eq22}) is valid for the physical case of reflecting Dirichlet
boundary conditions. See (\ref{Eq21}) for the case of reflecting
Neumann boundary conditions.

\bibitem{Notealp} The physical parameter $\alpha$ couples the massless scalar field
directly to the electromagnetic Maxwell tensor of the central
charged sphere [see Eqs. (\ref{Eq1}), (\ref{Eq2}), and (\ref{Eq3})].

\bibitem{Notedls} Note that (\ref{Eq22}), (\ref{Eq23}), and
(\ref{Eq24}) are dimensionless expressions.

\end{thebibliography}
\end{document}